\documentclass[12pt]{article}
\usepackage[latin1]{inputenc}

\usepackage{amsmath}
\usepackage{color}
\usepackage{amsfonts}
\usepackage{amssymb}
\usepackage[mathscr]{euscript}
\usepackage{graphicx}
\usepackage{geometry}
\usepackage{amssymb,epsfig,subfigure}
\usepackage{hyperref}
\usepackage{comment}
\usepackage{tabularx}
\usepackage{bm}
\usepackage{euscript}
\usepackage{graphicx}
\usepackage{color}
\usepackage{amsfonts}
\usepackage{exscale}
\usepackage{amsbsy}
\usepackage{subfigure}
\usepackage{textcomp}
\usepackage{comment}
\usepackage{hyperref}
\usepackage{slashed}
\usepackage{authblk}
\usepackage{tabularx}
\usepackage{euscript}
\usepackage{graphicx}
\usepackage{color}
\usepackage{amsfonts}
\usepackage{exscale}
\usepackage{amsbsy}
\usepackage{subfigure}
\usepackage{textcomp}
\usepackage{comment}
\usepackage{hyperref}
\usepackage{bm}
\usepackage{wrapfig}
\usepackage[font=footnotesize,labelfont=bf]{caption}

\def\ba#1\ea{\begin{align}#1\end{align}}
\def\bg#1\eg{\begin{gather}#1\end{gather}}
\def\bm#1\em{\begin{multline}#1\end{multline}}
\def\bmd#1\emd{\begin{multlined}#1\end{multlined}}

\newcommand{\be}{\begin{equation}}
\newcommand{\ee}{\end{equation}}
\newcommand{\bea}{\begin{eqnarray}}
\newcommand{\eea}{\end{eqnarray}}

\newcommand{\pd}{\partial}

\newcommand{\matleft}{\left(\begin{array}}
\newcommand{\matright}{\end{array}\right)}

\newcommand{\op}{\operatorname}

\usepackage[numbers,sort&compress]{natbib}
\def\simge{
    \mathrel{\rlap{\raise 0.511ex 
        \hbox{$>$}}{\lower 0.511ex \hbox{$\sim$}}}}

\def\simle{
    \mathrel{\rlap{\raise 0.511ex 
        \hbox{$<$}}{\lower 0.511ex \hbox{$\sim$}}}}

\makeatletter
\renewcommand\section{\@startsection {section}{1}{\z@}%
                                 {-3.5ex \@plus -1ex \@minus -.2ex}
                                   {2.3ex \@plus.2ex}%
                                   {\normalfont\large\bfseries}}
\renewcommand\subsection{\@startsection{subsection}{2}{\z@}%
                                   {-3.25ex\@plus -1ex \@minus -.2ex}%
                                     {1.5ex \@plus .2ex}%
                                     {\normalfont\bfseries}}
\renewcommand\subsubsection{\@startsection{subsubsection}{3}{\z@}%
                                   {-3.25ex\@plus -1ex \@minus -.2ex}%
                                     {1.5ex \@plus .2ex}%
                                     {\normalfont\itshape}}
\makeatother

\def\pplogo{\vbox{\kern-\headheight\kern -29pt
\halign{##&##\hfil\cr&{\ppnumber}\cr\rule{0pt}{2.5ex}&\ppdate\cr}}}
\makeatletter
\def\ps@firstpage{\ps@empty \def\@oddhead{\hss\pplogo}%
  \let\@evenhead\@oddhead 
}
\def\maketitle{\par
 \begingroup
 \def\thefootnote{\fnsymbol{footnote}}
 \def\@makefnmark{\hbox{$^{\@thefnmark}$\hss}}
 \if@twocolumn
 \twocolumn[\@maketitle]
 \else \newpage
 \global\@topnum\z@ \@maketitle \fi\thispagestyle{firstpage}\@thanks
 \endgroup
 \setcounter{footnote}{0}
 \let\maketitle\relax
 \let\@maketitle\relax
 \gdef\@thanks{}\gdef\@author{}\gdef\@title{}\let\thanks\relax}
\makeatother

\numberwithin{equation}{section}

\begin{document}

\setcounter{page}0
\def\ppnumber{\vbox{\baselineskip14pt
}}

\def\ppdate{
} \date{\today}

\title{\LARGE \bf Dirac Composite Fermions and Emergent Reflection Symmetry about Even Denominator Filling Fractions}
\author{Hart Goldman and Eduardo Fradkin}
\affil{ \it Department of Physics and Institute for Condensed Matter Theory,\\  \it University of Illinois at Urbana-Champaign, \\  \it 1110 West Green Street, Urbana, Illinois 61801-3080, USA}
\maketitle

\begin{abstract}
Motivated by the appearance of a ``reflection symmetry'' in transport experiments and the absence of statistical periodicity in relativistic quantum field theories, we propose a series of relativistic composite fermion theories for the compressible states appearing at filling fractions $\nu=1/2n$ in quantum Hall systems. These theories consist of electrically neutral Dirac fermions attached to $2n$ flux quanta via an emergent Chern-Simons gauge field. While not possessing an explicit particle-hole symmetry, these theories reproduce the known Jain sequence states proximate to $\nu=1/2n$, and we show that such states can be related by the observed reflection symmetry, at least at mean field level. We further argue that the lowest Landau level limit requires that the Dirac fermions be tuned to criticality, whether or not this symmetry extends to the compressible states themselves.
\end{abstract}

\bigskip
\newpage

\tableofcontents
\thispagestyle{empty}

\newpage
\setcounter{page}{1}

\section{Introduction}

A paradigmatic example of a strongly interacting metallic state arises in the context of 2d systems of electrons in a strong magnetic field when the lowest Landau level (LLL) is at filling $\nu=1/2$. However, despite many years of effort, concrete theoretical understanding of this state remains elusive. Historically, the most successful approach to this problem has been that of Halperin, Lee, and Read (HLR) \cite{halperin-1993}, which utilizes the notion of flux attachment, in which a theory of non-relativistic particles is exactly mapped to a theory of ``composite particles'' (fermions or bosons) coupled to an Abelian Chern-Simons gauge field \cite{Wilczek-1982}. Such mappings have been foundational in the theory of the fractional quantum Hall (FQH) effect \cite{Jain-1989,Zhang-1989,Lopez-1991}, explaining the observed Jain sequence FQH states as integer quantum Hall (IQH) states of composite fermions. In the HLR approach, two flux quanta are attached to each electron, completely screening the magnetic field and yielding a theory of a Fermi surface of non-relativistic composite fermions $f$, strongly coupled to a Chern-Simons gauge field $a_\mu=(a_t,a_x,a_y\,)$,
\be
\mathcal{L}_{\mathrm{HLR}}=f^\dagger (i\pd_t+\mu+a_t)f-\frac{1}{2m}|(i\pd_i+a_i+A_i\,)f|^2+\frac{1}{4\pi}\frac{1}{2}ada+\cdots\,,
\ee
where $A_i=\frac{B}{2}(x\hat{y}-y\hat{x})$ is the background vector potential, and we use the notation $AdB=\varepsilon^{\mu\nu\lambda}A_\mu\pd_\nu B_\lambda$. Here we require that the emergent gauge field cancels the external magnetic field, i.e. $\langle\varepsilon^{ij}\partial_i a_j\rangle=-\varepsilon^{ij}\pd_i A_j$. The HLR theory has seen great phenomenological success: it explains the existence of the observed metallic state \cite{Jiang1989}, and the large cyclotron radii of the composite fermions near $\nu=1/2$ lead to quantum oscillations which have been observed experimentally \cite{Kang1993,Smet1998,Smet1999,Willett1999,Kamburov2012}. 

Nevertheless, the HLR theory suffers from several well known problems. First, as a theory of a Fermi surface strongly coupled to a gauge field, it is plagued by infrared (IR) divergences, and the random phase approximation (RPA) is uncontrolled. Additionally, the HLR theory is not a proper LLL theory, since a theory composite fermions which are charged under electromagnetism will not have holomorphic wave functions \cite{Read1994}. Finally, while the LLL Hamiltonian at $\nu=1/2$ is particle-hole ($\mathbf{PH}$) symmetric \cite{Girvin1984}, and $\mathbf{PH}$ symmetric response has been observed experimentally \cite{Wong1996,Shahar1995,Shahar1996}, the HLR theory does not seem to possess this symmetry: flux is attached to electrons, rather than holes. This issue has also found new relevance in recent quantum oscillation experiments \cite{Kamburov2014,Deng2016}. 

A great deal of progress on the latter two problems was made recently, when Son proposed a Dirac composite fermion theory of the $\nu=1/2$ state \cite{Son2015}. This theory is based on the fact that the LLL limit of a system of electrons with gyromagnetic ratio $g=2$ can be identified with the massless limit of a Dirac ``electron'' in a magnetic field
\be
\label{eq: free Dirac}
\mathcal{L}_e=i\bar{\Psi}\slashed{D}_A\Psi+\frac{1}{8\pi}AdA\,,
\ee       
where we have introduced the notation $D_A^\mu=\pd^\mu-iA^\mu$ and $\slashed{D}=D^\mu\gamma_\mu$, and $\gamma_\mu$ are the Dirac gamma matrices. The term $AdA/8\pi$ can be thought of as coming about due to the presence of a heavy fermion doubler\footnote{Throughout this paper, we approximate the Atiyah-Patodi-Singer $\eta$-invariant as a level-$1/2$ Chern-Simons term and include it in the action.}. Since Dirac fermions have Landau levels with both positive and negative energies, with one sitting at zero energy, the zeroth Landau level is half filled when the chemical potential is zero. Such a state is automatically symmetric under $\mathbf{PH}$, which is just the exchange of empty and filled states. This led Son to conjecture that this theory is dual to one of Dirac composite fermion vortices $\psi$ at finite density, strongly coupled to an emergent gauge field $a_\mu$ without a Chern-Simons term (QED$_3$),
\be
\label{eq: Son's CDL theory}
\mathcal{L}_{\mathrm{Son}}=i\bar{\psi}\slashed{D}_a\psi+\frac{1}{4\pi}adA+\frac{1}{8\pi}AdA+\cdots\,.
\ee 
where the $\mathbf{PH}$ symmetry of the Dirac electron problem now manifests as a time reversal ($\mathbf{T}$) symmetry of the composite fermions. The $\cdots$ denote irrelevant operators, such as the Maxwell term for $a_\mu$. This duality between a free Dirac fermion and QED$_3$ was quickly shown to be a part of a ``web of dualities,'' at the center of which is a relativistic flux attachment duality relating a free Dirac fermion to a Wilson-Fisher boson coupled to a Chern-Simons gauge field \cite{Seiberg2016,Karch2016}. This fermion-vortex duality has also led to progress in other areas of condensed matter physics \cite{Wang2015,Metlitski2016,Goldman2017,Hui:2017cyz}. Despite its success in incorporating $\mathbf{PH}$ symmetry, it still remains to understand how Son's theory might emerge from microscopics and the extent to which it can be experimentally distinguished from the HLR theory, although very interesting arguments have been put forward suggesting that Son's theory may emerge from the HLR theory upon incorporating the effect of quenched disorder \cite{Wang2017,Kumar2018,Kumar2018a} or as a percolation transition between the HLR theory and its $\mathbf{PH}$ conjugate \cite{Mulligan2016}. Encouragingly, evidence for the Dirac composite fermion theory has been found in numerical studies \cite{Geraedts2015,Geraedts2017}.

A major open question has been whether Son's proposal can be extended to describe the compressible states appearing at other even denominator filling fractions $\nu=1/2n$. In the HLR theory, descriptions of these states arise trivially: one can simply attach an even number of flux quanta so that the external field is again completely screened. For non-relativistic particles, this transformation should be an identity at the level of the partition function, implying that all of these theories lie in the same universality class. On the other hand, in relativistic theories, flux attachment influences both statistics and spin, and so this transformation is no longer innocuous. More saliently, while the LLL Hamiltonian for these states is not $\mathbf{PH}$ symmetric, transport experiments have observed an analogous ``reflection symmetry'' in the $I-V$ curves about the $\nu=1/3$ FQH to insulator transition, which occurs at $\nu=1/4$ \cite{Shahar1995,Shahar1996,Shahar1997}, suggesting that this state might host its own kind of $\mathbf{PH}$ symmetry, or at least that there is a symmetry relating the Jain sequence FQH states proximate to it. More precisely, the observed symmetry maps a longitudinal $I-V_{xx}$ curve at a filling fraction $\nu<1/2n$ to a $I-V_{xx}$ curve at a dual filling fraction $\nu'>1/2n$ in which the roles of current and voltage are exchanged, 
\be
(E_i(\nu),J_i(\nu))=\left(\frac{h}{e^2}J_i(\nu'),\frac{e^2}{h}E_i(\nu')\right)\,. 
\ee
and the transverse $I-V_{xy}$ curves in the observed region are all linear with slope $3\frac{h}{e^2}$. Surprisingly, the observed longitudinal $I-V_{xx}$ curves do not appear to be linear except very close to $\nu=1/4$, meaning that the observed symmetry extends to \emph{nonlinear} response.  

The presence of this symmetry makes sense if we view the state at $\nu=1/2n$ as a limit of the Jain states $\nu=p/(2np+1)$, where $n$ and $p$ are integers. Such states have reflection conjugates on either side of the point at $\nu=1/2n$, and we thus might expect the composite fermions in these states to experience the same physics. Indeed, to impressively high precision, 
the reflection symmetry observed in experiment appears identical to the one which relates conjugate Jain states. However, the HLR theory is incompatible with this symmetry, since the conjugate states in question correspond to \emph{different} IQH states of composite fermions.

In this article, we propose a series of Dirac composite fermion theories to describe the compressible states at $\nu=1/2n$. These theories are obtained by attaching an even number of fluxes to the composite fermions of Son's theory \eqref{eq: Son's CDL theory}. They therefore can be thought of as existing in a unified framework with Son's theory of $\nu=1/2$. Although they lack an explicit analogue of $\mathbf{PH}$ symmetry, we argue that they can explain the reflection symmetry observed in experiments. In particular, we show that in reflection conjugate Jain states, the composite fermions fill the \emph{same} number of Landau levels, in contrast to HLR. Moreover, our theories are consistent with the LLL limit: as in Son's theory, our Dirac composite fermions are electrically neutral. 
In fact, we show that the LLL limit 
ensures that the Dirac composite fermion is massless, 
whether or not a mass is allowed by symmetry. 
If a mass is indeed allowed by symmetry, that would suggest that the states at $\nu=1/2n$ for $n>1$ are tuned to a quantum critical point, rather than constituting a genuine phase.

It is not clear to us whether the reflection symmetry of the Jain states proximate to $\nu=1/2n$ extends to a full blown symmetry of the theories \emph{at} $\nu=1/2n$ for $n>1$. The case of $\nu=1/2$ is special in this regard, since there the reflection symmetry is identical to $\mathbf{PH}$ symmetry, which manifests itself as the $\mathbf{T}$ symmetry of the composite fermion theory. While the experiments do strongly hint that the compressible states at $\nu=1/2n$ have this reflection symmetry, they do not necessarily imply it. This is because the experiments may not have truly observed the compressible state, instead seeing the signatures of 
the phases asymptotically close to $\nu=1/4$. However, it is entirely possible that our composite fermion theories flow to fixed points hosting an enhanced $\mathbf{T}$ symmetry which is the continuation of the reflection symmetry of the Jain states. The presence of such a symmetry would also ensure the masslessness of the Dirac fermions, and it would imply that our theories display ``self-dual" transport at $\nu=1/2n$ \cite{Shahar1997}. 


We finally note that other descriptions of the $\nu=1/2n$ states have been proposed in Refs. \cite{Wang2016,You2017} using semiclassical arguments\footnote{We thank Yizhi You for very enlightening discussions about these theories and their relationship to those presented in this paper.}. These theories are variants of HLR involving Fermi surfaces with nonvanishing Berry phases, which are related to an ``anomalous velocity" term associated with the non-commutative geometry of the LLL  \cite{Haldane2004}. The effect of these Berry phases is to generate anomalous Hall conductivities that completely cancel the Chern-Simons terms of HLR. However, it is not clear whether this cancellation truly occurs beyond $\nu=1/2$: without $\mathbf{PH}$ symmetry, the Berry phase can run. 
These theories also do not seem compatible with the reflection symmetry of the Jain states. Our expectation is that the same kind of anomalous velocity that is associated with the Berry phase 
can be equally well explained via interactions with a Chern-Simons gauge field
. Evidence for this comes from the fact that our theories lead to the same set of magnetoresistance minima as the theories of Fermi surfaces with $\pi/n$ Berry phases. However, it is difficult to make these connections precise because band theory intuition cannot be applied to our strongly interacting problem. In future work, we hope to elucidate the connections between these theories and the ones presented here.

We proceed as follows. In Section \ref{section: Lagrangian}, we present our proposed effective field theories for the $\nu=1/2n$ states. In Section \ref{section: reflection symmetry}, we describe how these theories can explain the reflection symmetry of the Jain states proximate to $\nu=1/2n$. In Section \ref{section: LLL limit}, we argue for the Dirac composite fermions should be massless by viewing the state at e.g. $\nu=1/4$ as the LLL limit of the HLR theory when the non-relativistic composite fermions are placed at half filling. We then discuss some additional observables in Section \ref{section: further checks}, in particular describing how to couple our theories to background geometry. We conclude in Section \ref{section: conclusion}.

\section{Proposed Effective Field Theories}
\label{section: Lagrangian}
We conjecture that the $\nu=1/2n$ state can be described as a theory of $2n$ flux quanta attached to a free Dirac fermion. This flux attachment transformation can be implemented on the Lagrangian \eqref{eq: free Dirac} by making the background gauge field dynamical, $A\rightarrow a$, and introducing a new auxiliary gauge field $c$ at level $2n$ that couples to $a$ and the background vector potential $A=\frac{B}{2}(x\hat{y}-y\hat{x})$ through BF terms. For a review of such flux attachment transformations, see Ref. \cite{Fradkin-2013}, 
\be
\label{eq: theory properly quantized}
i\bar{\psi}\slashed{D}_a\psi-\frac{1}{8\pi}ada+\frac{1}{2\pi}adc-\frac{2n}{4\pi}cdc+\frac{1}{2\pi}cdA\,.
\ee
We note that these transformations are the elements $\mathcal{S}\mathcal{T}^{-2n}\mathcal{S}$ of the modular group PSL$(2,\mathbb{Z})$, described in condensed matter and high energy contexts by Kivelson, Lee, and Zhang \cite{Kivelson1992} and Witten \cite{Witten2003} respectively. 
This theory is gauge invariant with all of the gauge fields satisfying the Dirac flux quantization condition. If we loosen this requirement (an innocuous thing if our interest is in local properties) and integrate out the auxiliary gauge field $c$, we arrive at the theory\footnote{In the remainder of this paper, we will work exclusively with theories having improperly quantized Chern-Simons levels.}
\be
\label{eq: theory}
\mathcal{L}_{1/2n}=i\bar{\psi}\slashed{D}_a\psi-\frac{1}{4\pi}\left(\frac{1}{2}-\frac{1}{2n}\right)ada+\frac{1}{2\pi}\frac{1}{2n}Ada+\frac{1}{4\pi}\frac{1}{2n}AdA\,.
\ee
notice that we recover Son's theory \eqref{eq: Son's CDL theory} for $n=1$. For $n>1$, this theory breaks $\mathbf{PH}$, $\mathbf{T}$, and parity ($\mathbf{P}$) due to the presence of the nonvanishing Chern-Simons term for $a$. Thus, na\"{i}vely, a Dirac mass is allowed by symmetry, unless this theory harbors an enhanced symmetry which prohibits a mass. In the absence of such a symmetry, these theories are taken to be tuned to a quantum critical point.  

The $\nu=1/2n$ state corresponds to the case where the composite fermions $\psi$ are at finite density but see a vanishing magnetic field. If we denote the physical electron density as $\rho_e=\left\langle\frac{\delta\mathcal{L}_{1/2n}}{\delta A_t}\right\rangle$ and the magnetic field seen by the composite fermions as $b_*=\langle\varepsilon^{ij}\pd_ia_j\rangle$, then 
\be
\label{eq: b_*}
\nu=2\pi\frac{\rho_e}{B}=\frac{1}{2n}\left(1+\frac{b_*}{B}\right)\,,
\ee
meaning that, indeed, $\nu=1/2n$ implies $b_*=0$. Thus, the composite fermions form a strongly interacting metallic state.

In the sections that follow, we will see that there are several reasons to believe that this theory correctly describes the physics of the $\nu=1/2n$ state. First, we will check that the IQH states of composite fermions reproduce  the Jain sequences, $\nu=p/(2np+1)$, where $p$ and $n$ are integers. We will then introduce a $\mathbf{PH}$-like reflection transformation which maps between Jain states on either side of $\nu=1/2n$, and we will see that conjugate Jain states correspond to \emph{the same} IQH state of composite fermions, up to a $\mathbf{T}$ transformation ($\nu\mapsto-\nu$). This transformation can be related to boson-vortex exchange upon invoking boson-fermion duality to obtain theories of composite bosons at $\nu=1$. This goes a long way toward explaining the reflection symmetry observed in experiments. 

The theories \eqref{eq: theory} are also consistent with the LLL limit. Not only is the composite fermion charge neutral, but we will use Son's particle-vortex duality to argue that the dual theory to Eq. \eqref{eq: theory}, given by
\be
\label{eq: dual theory}
\tilde{\mathcal{L}}_{1/2n}=i\bar{\chi}\slashed{D}_{b}\chi+\frac{1}{8\pi}bdb+\frac{1}{4\pi}\frac{1}{2(n-1)}(b+A)d(b+A)\,,
\ee 
where $b$ is another emergent gauge field (\emph{note the difference with $b_*$}), reproduces the same LLL physics as a theory of non-relativistic electrons with $2(n-1)$ flux quanta attached, at least at mean field level. This leads to an explanation for why $\psi$ and $\chi$ are massless, despite the fact that a mass may be allowed by symmetry. Moreover, it is easy to see that the $\nu=1/2n$ state corresponds to a half filled zeroth Landau level of $\chi$ particles.

\section{Reflection Symmetry of the Jain Sequences}
\label{section: reflection symmetry}
\subsection{Reproducing the Jain Sequences}

We now show that the theories \eqref{eq: theory} reproduce the Jain sequences. We will see that the presence of the Chern-Simons term is crucial in making this work out. For simplicity, we will work with the version of the theory with improperly quantized Chern-Simons levels, although the computations for the properly quantized theory are similar. 
We start by considering the equation of motion for $a_t$,
\be
0=\langle\psi^\dagger\psi\rangle-{1\over2\pi}\left({1\over2}-{1\over2n}\right)b_*+{1\over2\pi}{1\over 2n}B\,,
\ee
meaning, if we define the composite fermion filling fraction as $\nu_\psi=2\pi\rho_\psi/b_*$, where $\rho_\psi=\langle\psi^\dagger\psi\rangle$, then
\be
\label{eq: nu_psi}
\nu_\psi={1\over2}-{1\over2n}-{1\over 2n}{B\over b_*}\,.
\ee
Notice that for $n=1$, the first two terms on the right hand side cancel, 
as the density of composite fermions is proportional to the background magnetic field in that case. For $n\neq1$, the non-cancellation of the first two terms reflects the fact that the Chern-Simons level is non-vanishing: the density of the composite fermions depends on the external magnetic field \emph{and} the emergent magnetic field, $b_*$. 

To produce the Jain sequences, we first fill $p$ Landau levels of the composite fermions, leading to an incompressible integer quantum Hall state
\be
\nu_\psi=p+{1\over2}\,.
\ee
Eq. \eqref{eq: nu_psi} is now
\be
\label{eq: b* Jain}
2np+1=-{B\over b_*}\,.
\ee
We know $B/b_*$ in terms of the physical electron filling fraction $\nu$ from Eq. \eqref{eq: b_*}. Plugging this in, we have
\be
2np+1=-{1\over{2n\nu-1}}\,.
\ee
Solving for $\nu$, we finally obtain
\be
\label{eq: Jain sequence}
\nu={p\over{2np+1}}\,.
\ee
This is the Jain sequence.

\subsection{Reflection Symmetry}
\label{section: toward self-duality}

Having shown that the theories \eqref{eq: theory} reproduce the Jain sequences, our goal now is to determine if they can reproduce 
``reflection" symmetry indicated by experiments, which relates the (nonlinear) response of 
conjugate Jain states proximate to $\nu=1/2n$. 
For us, this amounts to showing that for each Jain state proximate $\nu=1/2n$, there is a conjugate Jain state where the composite fermions fill the same number of Landau levels but are $\mathbf{T}$ conjugated. At least at mean field level, the composite fermion response of such states should be essentially identical. 
Note that while this symmetry of the incompressible plateau states can give us intuition that there is an emergent reflection symmetry at the compressible state $\nu=1/2n$, such a symmetry is not implied. It is interesting enough that the theory \eqref{eq: theory} explains the symmetry of the plateau states.

Each state on the Jain sequence \eqref{eq: Jain sequence} with filling $\nu<1/2n$ ($p\geq0$) has a reflection conjugate with filling $\nu'>1/2n$ given by
\be
\label{eq: Jain conjugation}
\nu'=\frac{1+p}{2n(1+p)-1}\,.
\ee
This can be written as the following transformation of the filling fraction $\nu$,
\be
\label{eq: self-duality of filling}
\nu'=\frac{-(2n-1)\nu+1}{[1-(2n-1)^2]\nu+(2n-1)}\,.
\ee
Notice that for $n=1$, this relation is none other than the $\mathbf{PH}$ transformation $\nu'=1-\nu$.

The statement of the reflection symmetry between the conjugate Jain states is that they correspond to composite fermion IQH states with $\nu_\psi'=-\nu_\psi=-(p+1/2)$. In other words, an IQH state of composite fermions is mapped to \emph{the same} IQH state 
up to a $\mathbf{T}$ transformation (i.e. with magnetic field pointing in the opposite direction). To see that this is the case, we start by rewriting Eqs. \eqref{eq: b_*} and \eqref{eq: nu_psi} as a relation between Dirac and electron filling fractions,
\be
\label{eq: nu_psi to nu_e}
\nu_\psi-\frac{1}{2}=\frac{\nu}{1-2n\nu}\,.
\ee
Plugging Eq. \eqref{eq: Jain conjugation} into Eq. \eqref{eq: nu_psi to nu_e}, the dependence on the compressible state index $n$ cancels, and we obtain
\be
\label{eq: PH conjugate CF filling}
\nu_\psi'=-\left(p+\frac{1}{2}\right)\,.
\ee
Thus, the conjugate state can be thought of as filling $p$ Landau levels with the magnetic field pointing in the opposite direction!  
This is a related to the more general fact that reflection symmetry acts as $\mathbf{T}$ on the composite fermion filling fraction: Eq. \eqref{eq: nu_psi to nu_e} implies that mapping $\nu\mapsto\nu'$ is the same as $\mathbf{T}: \nu_\psi\mapsto -\nu_\psi$. Note that in the language of the dual theory, Eq. \eqref{eq: dual theory}, this $\mathbf{T}$ symmetry can be interpreted as a particle-hole symmetry $\mathbf{CT}$ (we reserve $\mathbf{PH}$ for the electron particle-hole symmetry $\nu\mapsto 1-\nu$). 

The above results go a long way toward explaining the reflection symmetry observed experimentally. However, it is important to note that since the composite fermion theories under consideration do not appear to be $\mathbf{T}$ symmetric at $\nu=1/2n$
, the physics of the Jain state at filling factor $\nu$ might differ from that at its conjugate $\nu'$ due to the effect of fluctuations of the emergent gauge field. This being said, since these states are gapped, we expect the effect of such fluctuations to be small and essentially unobservable, and we believe that this mean field argument should suffice to explain what is observed in experiments. As the compressible state is approached, however, gauge field fluctuations will become important, and the reflection symmetry may be broken. Whether the symmetry persists to the compressible state ultimately requires an understanding of the interplay of disorder and the strong interactions with the Chern-Simons gauge field. In the next subsection, we will consider the implications of this reflection symmetry for transport in more detail and discuss what the experimental observations can tell us about whether this symmetry emerges at the compressible states at $\nu=1/2n$. 

We can develop a complementary interpretation of the transformation \eqref{eq: self-duality of filling} in the language of composite bosons as an exchange symmetry between composite bosons and vortices, or ``self-duality''. A similar interpretation was also introduced in the previous, non-relativistic approaches to this problem \cite{Kivelson1992,Shahar1996,Shahar1997}. It will also be a particularly useful language for writing down constraints on transport, which is the topic of the next subsection. We can obtain a composite boson theory by invoking the duality between a gauged Wilson-Fisher boson and a free Dirac fermion described in Refs. \cite{Seiberg2016,Karch2016}, 
\be
i\bar{\Psi}\slashed{D}_A\Psi+\frac{1}{8\pi}AdA\longleftrightarrow|D_a\phi|^2-|\phi|^4-\frac{1}{4\pi}ada+\frac{1}{2\pi}adA\,,
\ee
where $\longleftrightarrow$ denotes duality and we use the notation ``$\,-|\phi|^4\,$'' to indicate tuning to the Wilson-Fisher fixed point. It is not difficult to see that the theories \eqref{eq: theory} have bosonic duals,
\be
\label{eq: composite bosons}
|D_{g-A}\phi|^2-|\phi|^4+\frac{1}{4\pi}\frac{1}{2n-1}g dg\,.
\ee
Here $g$ is another fluctuating emergent gauge field. For $A={B\over2}(x\hat{y}-y\hat{x})$, these bosons find themselves at finite density and magnetic field. Differentiating with respect to $g_t$ and $A_t$ gives
\be
\left\langle j^t_\phi\right\rangle=-\rho_e=-\frac{1}{2n-1}\frac{\langle \varepsilon^{ij}\pd_ig_j\rangle}{2\pi}\,,
\ee
where $j_\phi^\mu$ is the \emph{gauged} $U(1)$ current of the bosons. If we define the filling of the bosons to be $\nu_\phi=2\pi\frac{\langle j_\phi^t\rangle}{\langle \varepsilon^{ij}\pd_i(g_j-A_j)\rangle}$, then
\be
\nu_\phi=-[2n-1-\nu^{-1}]^{-1}\,,
\ee 
Thus, for $\nu=\frac{1}{2n}$, we have
\be
\nu_\phi=1\,.
\ee
Thus, we now are facing a problem of gauged Wilson-Fisher bosons at $\nu_\phi=1$. Interestingly, the filling of the bosons in this theory is independent of $n$.

The theory \eqref{eq: composite bosons} can be shown to be dual to a theory of bosonic vortices \cite{Peskin1978,Dasgupta1981} with action\footnote{For explicit derivations of the particular boson-vortex duality discussed here, see Refs. \cite{Goldman2018,Mross2017}.} 
\be
\label{eq: composite bosons - dual}
|D_h\tilde\phi|^2-|\tilde\phi|^4-\frac{2n-1}{4\pi}hdh+\frac{1}{2\pi}hdA\,,
\ee
where $g$ is a new emergent gauge field. Notice that, in the composite boson language, the lack of explicit $\mathbf{T}$ symmetry in the fermionic theory manifests itself as an apparent lack of symmetry between bosons and vortices: they are interacting with Chern-Simons gauge fields having different levels. Similar manipulations to those for the $\phi$ theory imply a relationship between the filling fractions of the dual theories,
\be
\nu_{\phi}=-\frac{1}{\nu_{\tilde\phi}}\,,
\ee
which is the modular $\mathcal{S}$ transformation of Refs. \cite{Kivelson1992,Witten2003}. Conjugate filling fractions are those for which the roles of bosons and vortices have been exchanged, i.e. $\nu_\phi(\nu)=-\nu_{\tilde\phi}(\nu')$. Thus,
\be 
\nu^{-1}-(2n-1)=\frac{1}{\nu'^{-1}-(2n-1)}\,.
\ee
Solving for $\nu'$ leads to Eq. \eqref{eq: PH conjugate CF filling}. Thus, reflection symmetry can be interpreted equivalently as a $\mathbf{T}$ (or $\mathbf{CT}$) symmetry of composite fermions and as a composite boson-vortex exchange symmetry, or ``self-duality.''

\subsection{Constraints on Transport and Connections to Experiment}


We now describe the implications of the reflection symmetry introduced above for transport and relate them to the experimental observations of Refs. \cite{Shahar1995,Shahar1996}. For convenience, we start by working with the composite boson description of Eqs. \eqref{eq: composite bosons} and \eqref{eq: composite bosons - dual} since the experiments are most easily interpreted in that language, although we  describe how to translate these results into the composite fermion language at the end of this section. Note that the arguments in this section are essentially the same as those for non-relativistic composite boson theories given in Ref. \cite{Shahar1997}. However, we emphasize that in our case we are starting with theories which naturally manifest the reflection symmetry of the Jain states proximate to $\nu=1/2n$ (as is evident in the Dirac composite fermion language). In the previous non-relativistic work, this symmetry had to be postulated.  

\subsubsection{Composite Boson Language: Self-Duality}

We start by defining the conductivity of the composite bosons of Eq. \eqref{eq: composite bosons}, which respond to both the background probe electric field $E_i$ and the emergent electric field due to the Chern-Simons gauge field $g$, $\langle e_i(g)\rangle=\langle f_{it}(g)\rangle$, 
\be
\label{eq: CB conductivity def}
\langle j_{\phi,i}\rangle=\sigma^{\mathrm{CB}}_{ij}(\langle e^j(g)\rangle-E^j)\,.
\ee
Note that, in this section, all conductivities (resistivities) are in units of $e^2/\hbar$ ($\hbar/e^2$).

In the composite vortex theory \eqref{eq: composite bosons - dual}, the roles of charge and flux are exchanged. Consequently, $\sigma_{ij}^{\mathrm{CB}}$ is the \emph{resistivity tensor} of the vortices. To see this, we plug the charge-flux mappings $j_\phi=dh/2\pi$ and $j_{\tilde\phi}=d(g-A)/2\pi$, i.e. $j_{\phi}^i=\varepsilon^{ij}(\pd_j h_t-\pd_t h_j)/2\pi\equiv\varepsilon^{ij}\tilde{e}_j/2\pi$ and $j_{\tilde{\phi}}^i=\varepsilon^{ij}(e_j-E_j)/2\pi$, into Eq. \eqref{eq: cb dictionary} to obtain the transport dictionary,
\be
\label{eq: cb dictionary}
\sigma^{\mathrm{CB}}_{ij}=\frac{1}{(2\pi)^2}\varepsilon^{ik}\varepsilon^{jl}\,\tilde{\rho}_{kl}^{\mathrm{CB}}
\,.
\ee
In a rotationally invariant system, this simply reduces to $\sigma^{\mathrm{CB}}_{ij}=\tilde{\rho}_{ij}^{\mathrm{CB}}/(2\pi)^2$. Because the dictionary \eqref{eq: cb dictionary} is a consequence of particle-vortex duality, it is valid at finite wave vector and frequency, as well as in the presence of disorder. Moreover, since we never explicitly required linear response in its derivation, we also expect the dictionary to hold beyond the linear regime. This last point is necessary if our wish is to understand the experiments of Refs. \cite{Shahar1995,Shahar1996}, since the symmetry observed there was one of nonlinear response.   

The equality between composite boson conductivity and vortex resistivity is the reason why the reflection symmetry described above exchanges the role of current and voltage about $\nu=1/2n$. In the bosonic language, the reflection symmetry is the statement that composite bosons at electron filling fraction $\nu$ have identical transport to the composite vortices at conjugate electron filling fraction $\nu'$, so
\be
\label{eq: reflection symmetry in transport}
\rho^{\mathrm{CB}}_{ij}(\nu)=\tilde{\rho}_{ji}^{\mathrm{CB}}(\nu')=(2\pi)^2\sigma_{ji}^{\mathrm{CB}}(\nu')
\,.
\ee
From the analysis of the Dirac composite fermion theories earlier in this section, we saw that this symmetry holds for FQH states proximate to $\nu=1/2n$, at least at mean field level.

We can connect $\rho^{\mathrm{CB}}_{ij}$ to the observable electron resistivity $\rho_{ij}$ as follows. If $J^i=-j_{\phi}^i$ is the electron current, then we define $\rho_{ij}$ via
\be
E_i=\rho_{ij}\langle J^j\rangle\,.
\ee 
The difference between $\rho_{ij}$ and $\rho^{\mathrm{CB}}_{ij}$ comes from a shift in the Hall resistivity due to the Chern-Simons gauge field, which enforces flux attachment, $\langle e_i\rangle=2\pi(2n-1)\varepsilon_{ij}\langle j^j_\phi\rangle$. Plugging this into the definition of $\sigma_{ij}^{\mathrm{CB}}$, \eqref{eq: CB conductivity def}, and rearranging, one finds
\be
\label{eq: rho rho_cb}
\rho_{ij}=\rho_{ij}^{\mathrm{CB}}-(2n-1)\,2\pi\varepsilon_{ij}\,.
\ee
Thus, the observed resistivity is just the composite boson resistivity with shifted Hall components.

If the reflection symmetry persists to $\nu=1/2n$ (which is mapped to itself), then Eq. \eqref{eq: reflection symmetry in transport} implies that the composite boson resistivity must satisfy the ``self-duality'' condition
\be
\label{eq: self-duality}
[\rho_{xx}^{\mathrm{CB}}(1/2n)]^2+[\rho_{xy}^{\mathrm{CB}}(1/2n)]^2=(2\pi)^2\,.
\ee   
For $n=1$, or $\nu=1/2$, this constraint and the relation \eqref{eq: rho rho_cb}, implies the $\mathbf{PH}$-symmetric Hall response $\sigma_{xy}=\frac{1}{4\pi}$ 
For $n\neq1$, however, the constraint is weaker: $\sigma_{xy}$ depends on the composite boson conductivity.  

We are now prepared to interpret the experimental results of Refs. \cite{Shahar1995,Shahar1996}, which correspond to the case of $\nu=1/4$, or $n=2$. Throughout the observed region of $\nu$ values, the Hall response was observed to be linear, with resistivity taking the value
\be
\label{eq: observed Hall resistivity}
\rho_{xy}=-3(2\pi)\,.
\ee  
meaning that, by Eq. \eqref{eq: rho rho_cb}, the Hall resistivity of the composite bosons vanishes
\be
\rho_{xy}^{\mathrm{CB}}=0\,.
\ee
This constraint is surprising, since \emph{there does not appear to be any symmetry in the problem which requires this.} Understanding the mechanism by which the composite boson Hall resistivity vanishes continues to be an open question. Plugging this into Eq. \eqref{eq: reflection symmetry in transport}, the reflection symmetry can be expressed in terms of the electron longitudinal resistivities as
\be
\label{eq: reflection of rhoxx}
\rho_{xx}(\nu)=\frac{(2\pi)^2}{\rho_{xx}(\nu')}\,.
\ee
since $\rho_{xx}^{\mathrm{CB}}=\rho_{xx}$. This is consistent with what was observed in the longitudinal $I-V_{xx}$ curves, assuming that Eqs. \eqref{eq: cb dictionary} and \eqref{eq: rho rho_cb} are valid in the nonlinear regime. It was also observed that, as $\nu$ approaches $1/4$, $\rho_{xx}$ seems to become linear and approach the ``self-dual'' value $\rho_{xx}=2\pi$. 
This constitutes fairly compelling evidence that reflection symmetry emerges at the compressible states at $\nu=1/2n$, but we emphasize that this is not necessary. In the next section, we will show that the LLL limit can suffice to tune the Dirac composite fermions to criticality, whether or not the states at $\nu=1/2n$ truly host an emergent reflection symmetry themselves.

\subsubsection{Composite Fermion Language: $\mathbf{T}$ Symmetry}

We close this section by considering the implications of reflection symmetry for the transport of the composite fermion theory \eqref{eq: theory}, for which the reflection symmetry is a $\mathbf{T}$ symmetry. If we define the composite fermion conductivity $\sigma^{\mathrm{CF}}_{ij}$ via
\be
\langle j_{\psi,i}\rangle=\sigma^{\mathrm{CF}}_{ij}\langle e^{j}(a)\rangle\,,
\ee
where $e_i(a)=f_{it}(a)$ and $j_\psi^\mu=\bar\psi\gamma^\mu\psi$, then reflection symmetry implies
\be
\label{eq: reflection as PH}
\sigma^{\mathrm{CF}}_{ij}(\nu)=\sigma^{\mathrm{CF}}_{ji}(\nu')\,.
\ee
Thus, if the reflection symmetry persists to $\nu=1/2n$, this means
\be
\sigma^{\mathrm{CF}}_{xy}=0\,.
\ee
Indeed, we will quickly see that the duality between the composite fermion theory \eqref{eq: theory} and the composite boson theory \eqref{eq: composite bosons} that this $\mathbf{T}$ symmetry implies self-duality of the bosons and vice versa. 

We can relate the composite fermion conductivity to the measured electron conductivity as follows. Differentiating the Lagrangian \eqref{eq: theory} with respect to $A_j$ and $a_j$ give the electron and composite fermion currents respectively
\be
\langle J_i\rangle=\frac{1}{2\pi}\frac{1}{2n}\varepsilon_{ij}(\langle e^j\rangle+E^j),\,\langle j_{\psi,i}\rangle=\frac{1}{2\pi}\left(\frac{1}{2}-\frac{1}{2n}\right)\varepsilon_{ij}\langle e^j\rangle-\frac{1}{2\pi}\frac{1}{2n}\varepsilon_{ij}E^j\,,
\ee
plugging in the definitions of the electron and composite fermion resistivities $\rho^{\mathrm{CF}}_{ij}=(\sigma^{\mathrm{CF}}_{ij})^{-1}$ (assuming rotation invariance) and solving the system of equations, one finds that the electron and composite fermion resistivities are related by
\bea
\rho_{xx}&=&(2\pi)^2\frac{4\rho^{\mathrm{CF}}_{xx}}{(\rho^{\mathrm{CF}}_{xx})^2+[\rho_{xy}^{\mathrm{CF}}+2(2\pi)]^2}\,,\\
\rho_{xy}&=&2\pi\left[-2(n-1)-8\pi\frac{2(2\pi)+\rho_{xy}^{\mathrm{CF}}}{(\rho^{\mathrm{CF}}_{xx})^2+[\rho_{xy}^{\mathrm{CF}}+2(2\pi)]^2}\right]\,.
\eea
There are several things to note about these expressions. First, reflection symmetry \eqref{eq: reflection as PH} along with the observed Hall resistivity $\rho_{xy}=-3(2\pi)$ again imply the observed $I-V_{xx}$ reflection symmetry, Eq. \eqref{eq: reflection of rhoxx}. Moreover, plugging $\mathbf{T}$ symmetry of the composite fermions ($\rho_{xy}^{\mathrm{CF}}=0$) into these equations and combining them with Eq. \eqref{eq: rho rho_cb} immediately leads to the self-duality of the composite bosons, Eq. \eqref{eq: self-duality}. Finally, assuming that $\mathbf{T}$ symmetry extends to the compressible state at $\nu=1/2n$ and plugging in the observed $\rho_{xy}=-3(2\pi)$ implies
\be
\sigma_{xx}^{\mathrm{CF}}(1/2n)=\frac{1}{4\pi}=\frac{1}{2}\frac{e^2}{h}\,.
\ee
Thus, the problem of understanding the physical origin of the observed Hall resistivity at the $\nu=1/3$ -- insulator transition may in fact be identical to the problem of understanding why $\sigma_{xx}^{\mathrm{CF}}=1/4\pi$. Intriguingly, this value is the same as that obtained in Pruisken's two-parameter scaling theory of IQH plateau transitions \cite{Pruisken1984,Pruisken1985}.

\section{Massless Composite Fermions from the LLL Limit}
\label{section: LLL limit}

We now argue that the LLL limit requires that our theories \eqref{eq: theory} be tuned so that the Dirac composite fermions are massless. This argument essentially follows the logic of Son's argument that the LLL limit of non-relativistic fermions with $g=2$ can be identified with that of a massless Dirac fermion \cite{Son2015}, provided that the effect of transitions between Landau levels is neglected. The difference here will be that instead of starting with a non-interacting theory of non-relativistic fermions in a magnetic field, we consider non-relativistic fermions in a magnetic field with $2(n-1)$ flux quanta attached via a Chern-Simons gauge field,
\be
if^\dagger D_{a,t}f-\frac{1}{2m}|D_{a,i}f|^2+\frac{\varepsilon^{ij}\pd_ia_j}{2m}f^\dagger f+\frac{1}{4\pi}\frac{1}{2(n-1)}(a+A)d(a+A)\,,
\ee
where again $A_i=\frac{B}{2}(x\hat{y}-y\hat{x})$ is the magnetic vector potential, and $A_t=\mu$. Here we work in a regime where $b=\langle\varepsilon_{ij}\pd_ia_j\rangle\neq0$, i.e. the fermions experience a net (uniform) magnetic field, organizing themselves into Landau levels. Our ultimate interest will be in the case where the non-relativistic composite fermions are at half-filling, which corresponds to a physical electron filling fraction $\nu=1/2n$.

The LLL limit of this theory $m\rightarrow0$ can be understood to be finite by introducing a Hubbard-Stratonovich field $c$ as follows,
\be
\label{eq: NR LLL}
if^\dagger D_{a,t}f+ic^\dagger(D_{a,x}+iD_{a,y})f-if^\dagger(D_{a,x}-iD_{a,y})c+2mc^\dagger c+\frac{1}{4\pi}\frac{1}{2(n-1)}(a+A)d(a+A)\,,
\ee
Upon taking the limit $m\rightarrow0$, we see that $c$ becomes a Lagrange multiplier which implements the LLL constraint $(D_{a,x}+iD_{a,y})f=0$.

We now argue that the theory \eqref{eq: NR LLL} is identical to that which would be obtained by taking the LLL limit of a Dirac fermion coupled to a Chern-Simons gauge field with Lagrangian, 
\be
\label{eq: Dirac LLL}
i\bar{\chi}\slashed{D}_{a}\chi+\frac{1}{8\pi}ada+\frac{1}{4\pi}\frac{1}{2(n-1)}(a+A)d(a+A)\,.
\ee
Notice that this theory is none other than the particle-vortex dual of our composite fermion theory for the $\nu=1/2n$ state \eqref{eq: theory}. Writing $\chi=(f,c)$ and choosing $\gamma^t=\sigma^z,\gamma^x=i\sigma^y,\gamma^y=-i\sigma^x$, we obtain
\be
if^\dagger D_{a,t}f+ic^\dagger D_{a,t}c+ic^\dagger(D_{a,x}+iD_{a,y})f-if^\dagger(D_{a,x}-iD_{a,y})c+\frac{1}{8\pi}ada+\frac{1}{4\pi}\frac{1}{2(n-1)}(a+A)d(a+A)\,.
\ee
This looks almost identical to our non-relativistic Lagrangian \eqref{eq: NR LLL}, with two differences. The first is the presence of a time derivative term for $c$. However, this term is negligible upon taking the LLL limit, which for a Dirac fermion is the limit\footnote{Here we have written the theory with Dirac fermion velocity $v=1$ (in units of the speed of light; $v$ is not to be confused with the Fermi velocity, $v_F=\pd_k\epsilon(k)|_{k_F}$, where $\epsilon(k)$ is the dispersion). Reintroducing $v$ and rescaling $c\mapsto c'=vc$, one sees that the term in question becomes $ic'^{\dagger} D_{t,a}c'/v^2$, which vanishes as $v\rightarrow\infty$.} of infinite fermion velocity, $v\rightarrow\infty$.  The second difference is the appearance of the parity anomaly term $ada/8\pi$, which can be thought of as implementing the effect of the Dirac sea of filled negative energy states. At mean field level, where $a$ is not dynamical, then this term would simply lead to a constant shift in the filling fraction and Hall conductivity with respect to the non-relativistic case, i.e.
\be
\nu_f=\nu_\chi+\frac{1}{2}\,.
\ee
However, the equivalence between the LLL physics of the Dirac and non-relativistic theories may be spoiled upon taking into account fluctuations of $a$, 
although it is reasonable to expect that fluctuations of $a$ about its mean field value do not contribute large corrections. 

Thus, the LLL limits of the non-relativistic, flux attached theory \eqref{eq: NR LLL} and the \emph{massless} Dirac fermion theory \eqref{eq: Dirac LLL} match, at least at mean field level, meaning that a proper description of the LLL requires tuning \eqref{eq: Dirac LLL} to criticality. In other words, if we view the problem of non-relativistic electrons at filling $\nu=1/2n$ as the $\nu=1/2$ state of of non-relativistic electrons attached to $2(n-1)$ units of flux (which should be an exact rewriting of the original problem), then the LLL limit connects the problem to one of massless Dirac fermions coupled to a Chern-Simons gauge field with its zeroth Landau level half filled, Eq. \eqref{eq: Dirac LLL}. We then obtain our Dirac composite fermion theory \eqref{eq: theory} upon invoking particle-vortex duality. The beauty of this approach is that we can leverage the flux attachment invariance of the underlying non-relativistic problem to obtain a relativistic composite fermion description of the states at $\nu=1/2n$, even though relativistic theories are not invariant under flux attachment (for an extended discussion of this point, see Ref. \cite{Goldman2018}). 

The analysis of this section leads to an interesting interpretation of the theories \eqref{eq: theory}. Unlike in HLR, where e.g. the composite fermion for the $\nu=1/2n$ state is related to that of the $\nu=1/2(n-1)$ state by attachment of two flux quanta, here the Dirac composite fermion of the $\nu=1/2n$ state is the \emph{dual vortex} of that at $\nu=1/2(n-1)$, placed at filling $3/2$. The reason the filling is $3/2$ instead of $1/2$ is related to the fact that the composite fermion Lagrangian \eqref{eq: theory} and its dual \eqref{eq: Dirac LLL} differ by a filled Landau level, $\frac{1}{4\pi}ada$. This actually makes sense from the perspective of Son's original duality, in which the state at $\nu=1/4$ is the $\nu=3/2$ state of the composite fermions.

\section{Further Observables}
\label{section: further checks}
\subsection{Shift and Hall Viscosity}

\subsubsection{Jain States}
We now describe how to couple our Dirac composite fermion theories \eqref{eq: theory} to background geometry and show that that it is possible to reproduce the remaining universal data associated with the Jain states: the total orbital spin per particle $s$ \cite{Wen1992,Wen1995}, which determines the shift of the Jain states on the sphere $\mathscr{S}=2s$, as well as the Hall viscosity \cite{Read2011},
\be
\eta_H=\frac{s\rho_e}{2}\,.
\ee
The Hall viscosity measures the response to external shear deformations, and is associated with stress tensor correlation functions. In Galilean invariant systems, it also determines the leading contribution to the Hall conductivity at finite wave vector \cite{Hoyos2011,Bradlyn2012}. 

In order to obtain $s$, we need to understand how to couple our Dirac composite fermions to the \emph{Abelian}\footnote{The presence of an Abelian spin connection breaks Lorentz invariance, but this is not problematic here since Lorentz invariance is already broken explicitly by the external magnetic field.} spin connection $\omega_\mu$.  
The strength of this coupling is the orbital spin of the Dirac composite fermions, $S_{z}$, which is \emph{not} restricted to be $1/2$. This is because flux attachment generally leads to a Berry phase which depends on the geometry. The presence of this Berry phase means that composite particles behave like they have an emergent ``fractional spin'' due to their strong interactions with the Chern-Simons gauge field\footnote{This can be thought of as a manifestation of the framing anomaly \cite{Witten1989}.}. 
The orbital spin of the composite fermions, $S_z$, which determines the coupling to $\omega_\mu$, can be identified with this fractional spin 
\cite{Cho2014}. This can be seen explicitly by rewriting the partition function of the theory \eqref{eq: theory} as a path integral over 
composite fermion worldlines, as described in detail in Ref. \cite{Goldman2018}. For the sake of brevity, however, we instead argue for the value of $S_z$ by analogy with the non-relativistic case, where attaching $2n$ flux quanta to a spinless fermion leads to an orbital spin $S_z=-n$ (the sign flip comes from integrating out the Chern-Simons gauge field). Since the Dirac fermion starts with spin $1/2$, we expect 
\be
\label{eq: fractional spin}
S_z=\frac{1}{2}-n\,.
\ee
We thus claim that our Dirac composite fermion theories can be coupled to geometry by using the covariant derivative $D^\mu(a,\omega)=\pd^\mu-ia^\mu+\frac{i}{2}\gamma^0\omega^\mu$ and shifting $A\rightarrow A+(2n-1)\omega/2$,
\begin{align}
\label{eq: theory with geometry}
\mathcal{L}_{1/2n}[\psi,a,A,\omega]=i\,&\bar\psi\,\gamma^\mu\left(\pd_\mu-ia_\mu+\frac{i}{2}\gamma^0\omega_\mu\right)\psi-\frac{1}{4\pi}\left(\frac{1}{2}-\frac{1}{2n}\right)ada\\
&-\frac{1}{2\pi}\frac{1}{2n}\left(A+{{2n-1}\over2}\omega\right)da+\frac{1}{4\pi}\frac{1}{2n}\left(A+{{2n-1}\over2}\omega\right)d\left(A+{{2n-1}\over2}\omega\right)\nonumber\\
&+\cdots\nonumber
\end{align}
where the $\cdots$ refer to additional purely gravitational contact terms which we will neglect and which are discussed in detail in Ref. \cite{Gromov2015}. Note that in this section we conjugate the sign of the BF term relative to the rest of the paper so that the Wen-Zee terms we obtain have positive sign.

Equipped with the Lagrangian \eqref{eq: theory with geometry}, we now proceed to calculate the shift of the Jain states on the sphere, from which we can extract the orbital spin $s$ as the coefficient of the Wen-Zee ($\frac{1}{2\pi}Ad\omega$) term when all of the dynamical fields have been integrated out. The degeneracy of the $p^{\mathrm{th}}$ Dirac fermion Landau level on the sphere is 
\be
d_p=\int d^2\mathbf{x}\,\frac{b_*}{2\pi}+2|p|\equiv N_\phi+2|p|\,.
\ee
This means that the number of composite fermions required to fill up to the $p^{\mathrm{th}}$ Landau level is
\be
N_{\mathrm{\psi}}
=N_\phi\left(p+\frac{1}{2}\right)+p\,(p+1)\,.
\ee
The shift $\mathscr{S}$ of the electron filling fractions on the Jain sequence is defined via
\be
N_e=\nu_e(N_\phi+\mathscr{S})\,.
\ee
To calculate $\mathscr{S}$, we start by integrating out the composite fermions. This generates new Chern-Simons and Wen-Zee terms, which are (for $b_*>0$)
\be
\frac{p+\frac{1}{2}}{4\pi}ada+\frac{p\,(p+1)}{4\pi}ad\omega\,.
\ee
So now we have a Lagrangian
\be
\frac{1}{4\pi}\left(p+\frac{1}{2n}\right)ada+\frac{1}{4\pi}\left[p(p+1)-\frac{2n-1}{2n}\right]ad\omega-\frac{1}{2\pi}\frac{1}{2n}adA+\frac{1}{4\pi}\frac{2n-1}{2n}Ad\omega+\frac{1}{4\pi}\frac{1}{2n}AdA+\cdots
\ee
Notice that the contribution of the parity anomaly of the Dirac composite fermion has been cancelled, so we can already expect that this will yield \emph{the same} answer that we would have obtained from HLR. We now integrate out $a$, which has the equation of motion,
\be
da=\frac{1}{2np+1}\left[dA+\frac{1}{2}\left(2n-1-2np(p+1)\right)d\omega\right]\,.
\ee
Thus, suppressing purely gravitational terms, we obtain
\be 
\frac{1}{4\pi}\frac{p}{2np+1}Ad\left(A+(p+2n)\omega\right)=\frac{\nu}{4\pi}Ad(A+\mathscr{S}\omega)\,.
\ee
Thus, the shift is  
\be
\mathscr{S}=p+2n\,,
\ee
which is precisely the known result for the Jain states \cite{Wen1995,Cho2014}! The orbital spin $s$ and Hall viscosity $\eta_H$ are therefore
\be
\label{eq: spin and Hall viscosity}
s=\frac{\mathscr{S}}{2}=\frac{p}{2}+n,\,\eta_{H}=\frac{1}{2}\left(\frac{p}{2}+n\right)\rho_e\,,
\ee
again consistent with previously known results.

\subsubsection{Some Speculation about the Compressible States}

We take this opportunity to speculate about the geometric response of the theories \eqref{eq: theory} at the compressible filling fractions $\nu=1/2n$. In the case of the state at $\nu=1/2$, this has been seen as a source of disagreement between Son's Dirac composite fermion theory and HLR, essentially because the composite fermion appears to have different orbital spin in the two approaches \cite{Levin2017}. Moreover, it is not even clear if the Hall viscosity should be viewed as universal in the HLR theory \cite{You2016}. However, recent  results seem to indicate consistency between the Dirac composite fermion approach and a non-relativistic ``bimetric theory'' of FQH states near $\nu=1/2$ \cite{Gromov2018}, and it appears likely that an approach starting from HLR with quenched disorder can match these results as well \cite{Wang2017,Kumar2018,Kumar2018a}.

A na\"{i}ve approach to obtaining the Hall viscosity for the compressible states might involve considering the result for the Jain states and taking the limit $p\rightarrow\infty$. Unfortunately, this clearly leads to a divergent result given Eq. \eqref{eq: spin and Hall viscosity}. However, we have already argued that the fractional spin of the composite Dirac fermions is given by \eqref{eq: fractional spin}. Since the composite fermions feel a vanishing magnetic field, this should be the only contribution to the total orbital spin $s$. Thus, we expect at mean field level,
\be
\label{eq: Hall viscosity at 1/2n}
\eta_H(\nu=1/2n)=-\frac{1}{2}S_z\rho_e=\frac{1}{2}\left(n-\frac{1}{2}\right)\rho_e\,.
\ee
This result should not receive large quantum corrections so long as the composite fermions remain massless, which we argued can be guaranteed both by the LLL limit as well as the reflection symmetry (assuming that it can be continued to $\nu=1/2n$). As already mentioned above, this quantity can be measured \cite{Delacretaz2017} from the finite wave vector part of the Hall response \cite{Hoyos2011,Bradlyn2012}, and so it may be possible to use this to distinguish our theories from HLR as well as the HLR-like theories with $\pi/n$ Berry phase discussed in the Introduction. 

\subsection{Quantum Oscillations}

Classic signatures of composite fermions are the quantum oscillations in magnetoresistance which occur as the filling is tuned away from $\nu=1/2n$, meaning that the composite fermions feel a small magnetic field. It is known that magnetoresistance minima occur along the Jain sequences $\nu=\frac{p}{2np+1}$, where the composite fermions feel a magnetic field $b_*$ which can be obtained from Eq. \eqref{eq: b* Jain},
\be
\frac{1}{b_*}=-\frac{p+\frac{1}{2n}}{\frac{B}{2n}}\,.
\ee
Up to the overall sign (which comes from the sign of the BF term in Eq. \eqref{eq: theory} and is a matter of convention), this is precisely the same result that would have been obtained from a theory of a Fermi surface with $\pi/n$ Berry phase, as in Refs. \cite{Wang2016,You2017}. In those references, the shift from an integer value in the numerator was seen as a consequence of the Berry phase. However, this shift can equally well be obtained by attaching flux to Dirac fermions.

\section{Discussion}
\label{section: conclusion}


In this article, we have proposed a series of Dirac composite fermion theories to describe the metallic states appearing at filling fraction $\nu=1/2n$ in quantum Hall systems. These theories are related to Son's theory of $\nu=1/2$ by attachment of $2n$ flux quanta. A major advantage of our theories is that they explain the $\mathbf{PH}$-like reflection symmetry observed in transport experiments, which relates Jain sequence states on either side of $\nu=1/2n$, since the composite fermions at conjugate filling fractions experience the same physics. No other theory presented thus far has been shown to accommodate these observations. In addition, we showed that at mean field level our theories are consistent with the LLL limit, provided that we view the state at e.g. $\nu=1/4$ as a half filled Landau level of the (non-relativistic) composite fermions at $\nu=1/2$. 

Many open questions remain. Foremost is the question of whether the reflection symmetry emerges at the compressible states at $\nu=1/2n$, rather than just being a property of their proximate phases. Answering this question conclusively from a theoretical point of view requires an understanding of the interplay of disorder and strong interactions in the Chern-Simons-matter theories we have presented here: even if reflection symmetry does not emerge in the clean limit, it may appear when disorder is introduced. Such problems are poorly understood at charge neutrality, let alone in the presence of a Fermi surface. Moreover, a potentially related issue is the problem of explaining the observed Hall resistivity at the $\nu=1/3$ FQH -- insulator transition \eqref{eq: observed Hall resistivity} (and also the $\nu=1$ IQH -- insulator transition), which does not appear to be set by any symmetry of the problem.  Progress on both of these issues can be made by studying the (uncontrolled) mean field problem with disorder \cite{Wang2017,Kumar2018,Kumar2018a}, exploiting new or existing dualities \cite{Goldman2017}, or by searching for perturbative approaches which can capture the effects of both disorder and interactions -- a direction which has been fruitful at least in the zero density limit and which can give us hints about general principles that can extend beyond the perturbative regime \cite{Goswami2017,Thomson2017}.  We intend to pursue all of these directions in the future.

It also remains to understand the precise relationship between the theories presented here and the theories of Fermi surfaces with $\pi/n$ Berry phases coupled to gauge fields (with no Chern-Simons term) introduced in Refs. \cite{Wang2016,You2017}. These theories are argued to emerge as a result of the non-commutative guiding center geometry of the LLL. However, as mentioned in the Introduction, it seems likely that our theories are also consistent with the geometry of the LLL, with the Chern-Simons term playing a similar role to the Berry phase. This is borne out by the fact that observables which na\"{i}vely appear to probe the Fermi surface Berry phase are the same in our theory. For example, quantum oscillation minima for the two theories are identical, and, in a $\nu=1/2n$ and $\nu=1-1/2n$ bilayer system, we expect that the $\pi$ Berry phase of our Dirac composite fermions should lead to the same suppression of $2k_F$ backscattering as seen in the $\pi/n$ Berry phase theories. Of course, a major distinguishing feature of our theories from the Berry phase theories is the reflection symmetry of the Jain sequence states.


{\it{Note added}} -- After this manuscript was  prepared, we became aware of new work by Jie Wang \cite{Wang2018}, which overlaps somewhat with our work and considers the same types of theories from a different but complementary perspective. 

\section*{Acknowledgements}
We thank Jing-Yuan Chen, Gil Young Cho, Thomas Faulkner, Yin-Chen He, Prashant Kumar, Sung-Sik Lee, Michael Mulligan, Srinivas Raghu, Dam Thanh Son, Jun Ho Son, and Yizhi You for helpful discussions and comments on the manuscript. This work was supported in part by the National Science Foundation (NSF) Graduate Research Fellowship Program under Grant No. DGE-1144245 (HG) and by the NSF under Grant No. DMR-1725401 (EF). HG is grateful for the support and hospitality of the Perimeter Institute for Theoretical Physics, where some of this work was carried out. Research at the Perimeter Institute is supported by the Government of Canada through the Department of Innovation, Science and Economic Development and by the Province of Ontario through
the Ministry of Research and Innovation.


%

\end{document}